\documentclass[epj,referee]{svjour}
\usepackage{amssymb}
\usepackage{graphics}
\begin{document}
\title{Diblock copolymer thin films: Parallel and perpendicular
lamellar phases in the weak segregation limit}
\author{Yoav Tsori \and David Andelman} 
\institute{School of Physics and Astronomy\\
  Raymond and Beverly Sackler Faculty of Exact Sciences\\
  Tel Aviv University, 69978 Ramat Aviv, Israel}
\date{Received: date / Revised version: date}
%
\abstract{We study morphologies of thin-film diblock copolymers
between two flat and parallel walls. The study is restricted to
the weak segregation regime below the order-disorder transition
temperature. The deviation from perfect lamellar shape is
calculated for phases which are perpendicular and parallel to the
walls. We examine the undulations of the inter material dividing
surface and its angle with the walls, and find that the deviation
from its unperturbed position
can be much larger than in the strong segregation case.
Evaluating the weak segregation stability of the lamellar phases,
it is shown that a surface interaction, which is quadratic in the
monomer concentration, favors the perpendicular lamellar phase. In
particular, the degeneracy between perpendicular and unfrustrated
parallel lamellar phases for walls without a preferential
adsorption is removed.
}
%
\authorrunning{Y. Tsori and D. Andelman}
\titlerunning{Diblock copolymer thin films:
Parallel and perpendicular ...}
\maketitle
\section{Introduction}\label{intro}
Diblock copolymers (BCP) are made up of two chemically distinct
chains covalently bonded together. The BCP system forms self
assembled structures with length scales in the nanometer to
micrometer range. On the level of mean-field theory, the bulk
phase diagram is governed by two parameters: $f=N_A/N$, the
fraction of the A-block in a chain of polymerization index
$N=N_A+N_B$, and $\chi N$, where $\chi$ is the Flory parameter
measuring the interaction between the two species, and is
inversely proportional to the temperature
\cite{B-F90,O-K86,M-SPRL94,M-B96}.

For temperatures above the order-disorder transition (ODT)
temperature the  system is in the disordered phase. As the
temperature is lowered, symmetric BCP melts ($f=\frac{1}{2}$)
undergo a weak first order transition to a lamellar phase at
$\chi>\chi_c$. As the degree of block asymmetry $f$ is increased,
$|f-\frac{1}{2}|>0$, other phases of hexagonal and cubic spatial
symmetries become stable  \cite{Semenov85,Leibler80,F-H87}.

The interfacial behavior of BCP melts has been the subject of
experimental \cite{L-RPRL94,K-WPRL96,M-RPRL97,W-RMM94}  and
theoretical
\cite{Fredrickson87,M-MPRE96,P-BMM97,matsenJCP97,G-M-B99,mm01,epl01,tilt01,turnerPRL92}
investigations. In the former case the substrate is typically
spin-coated by the BCP, and subsequently analyzed by small angle
neutron and X-ray scattering or neutron reflectivity measurements.
If the walls are neutral, i.e., without preferential adsorption
to one of the two blocks, thin films of lamellar diblock
copolymers maintain their bulk periodicity $d_0$ by aligning
perpendicular to the confining walls. Such long range ordering
can be transferred by various techniques to a surface, creating a
template useful in nanolithography \cite{Chaikin97}. In cases
where the walls prefer one of the two blocks, the lamellae can
reduce the interfacial interactions by aligning parallel to the
walls, and change the lamellar periodicity from its bulk value
$d_0$. Which of the two phases prevails (parallel or
perpendicular) depends on the distance $2L$ between the two walls
(film thickness), the strength of the wall interactions as well
as the degree of segregation $N\chi$.

Numerical calculations of confined BCP have been performed using
self-consistent field theory \cite{matsenJCP97,G-M-B99} and
Monte-Carlo simulations \cite{G-M-B99,wang00}. Using these
techniques, order parameter profiles and phase diagrams have been
obtained. Previous analytic theories \cite{wang00,pwMM,pwPRL},
while providingvaluable qualitative results, have been sensitive
to the specific choice of phenomenological coefficients, and this
sensitivity leads to marked inaccuracies of the order parameter
as compared to Monte-Carlo simulations \cite{wang00}.

In the present work we complement the numerical studies by
introducing an alternative  analytical method. In particular, we
derive the deviation of the perpendicular and parallel lamellae
from their bulk shape. In Sec. \ref{model} we introduce a model
free energy and derive the underlying equations. In Sec.
\ref{profiles} the shape of confined lamellae is investigated and
found to be, in general, very different from the bulk shape. The
energy of the perpendicular and parallel lamellae as a function
of surface separation $2L$ as well as the stability diagram is
discussed in Sec. \ref{energy}.

\section{Model}\label{model}

Close to the phase transition point (ODT) between the disordered
and lamellar phases, the free energy of symmetric BCP melt is
well described by the following Ginzburg-Landau expansion
\cite{F-H87,mm01,epl01,binder97,SH77,C-C98}:

\begin{equation}\label{Fb}
{\cal
F}_b=\int\left\{\frac12\tau\phi^2+\frac12h\left(\nabla^2\phi+
q_0^2\phi\right)^2 +\frac{u}{4!}\phi^4 \right\}{\rm
d}^3{\textbf r}
\end{equation}
The bulk free energy ${\cal F}_b$ (in units of $k_BT$) is given
as a functional of the local order parameter $\phi({\textbf
r})\equiv\phi_A({\textbf r})-f$, which is the deviation of the A
monomer concentration from its average value. The parameters
above are given by
\begin{eqnarray}
f&=&1/2\nonumber\\
q_0&\simeq& 1.95/R_g~~;~~\tau=2\rho N\left(\chi_c-\chi\right)
\label{consts1}
\end{eqnarray}
With a monomer size $a$, the gyration radius for
Gaussian chains is $R_g^2\simeq \frac16Na^2$, and $\rho=1/Na^3$. Other parameters in Eq. (\ref{Fb}) are
\begin{equation}
\chi_c\simeq 10.49/N~~;~~h=3\rho c^2 R_g^2/2q_0^2\label{consts2}
\end{equation}

The dimensionless parameters $u/\rho$ and $c$ are of order unity.
The reduced temperature $\tau\sim(\chi_c-\chi)$ is positive in
the disordered phase, where $\phi({\textbf r})=0$. Close to the
ODT the bulk system is described by two length scales: the first
is the periodicity of lamellar modulations $d_0=2\pi/q_0$, and
the second is the correlation length $\sim(\tau/h)^{-1/4}$,
characterizing the decay of surface induced modulations. This
length diverges at the ODT, $\chi=\chi_c$.

The interaction free energy of a BCP melt with the confining wall
(in units of $k_BT$) can be written as a sum of two terms
\begin{equation}\label{Fs}
{\cal F}_s=\int\left[\sigma({\bf r}_s)\phi({\bf
r}_s)+\tau_s\phi^2({\bf r}_s)\right]{\rm d^2{\bf r}_s}
\end{equation}
where $\{ {\bf r}_s\}$ denotes the wall position. The first term
is linear in the order parameter, and expresses preferential
adsorption: a positive $\sigma({\bf r}_s)$ induces a negative
$\phi({\bf r}_s)$ (preference to the B monomers). The second
(quadratic) term allows surface deviation of the Flory parameter
$\chi$ from its bulk value. A positive $\tau_s$ means that the
surface has an ordering temperature lower than the bulk one
\cite{lai}.

In the following we consider a thin film in which the melt is
confined by two flat and parallel walls at $y=\pm L$.
Interactions between the wall and the melt are assumed to be
short-range, and for homogeneous walls, $\sigma({\bf
r}_s)=const.$, used throughout this paper, no additional surface
length scales are introduced (see Refs.
\cite{mm01,epl01,tilt01,muthu97} where $\sigma({\bf r}_s)$ varies
on the walls). The strength of wall interaction is given by two
parameters: $\sigma^+=\sigma(y=L)$ and $\sigma^-=\sigma(y=-L)$.
Symmetric ($\sigma^+=\sigma^-$) and asymmetric
($\sigma^+=-\sigma^-$) walls will be considered as special cases.

The deviation of the order parameter, $\phi({\bf r})$, from its
bulk value $\phi_b({\bf r})$ is denoted by $\delta \phi$
\begin{equation}\label{def_del}
\delta\phi({\bf r})\equiv\phi({\bf r})-\phi_b({\bf r})
\end{equation}
This deviation contains the effect of the walls. The free energy
${\cal F}={\cal F}_b+{\cal F}_s$ is then expanded to second order
around its bulk value, ${\cal F}={\cal F}[\phi_b]+\Delta {\cal
F}[\delta\phi,\phi_b]$,
\begin{eqnarray}\label{deltaF}
\Delta {\cal F} &=&\int \left\{[(\tau+hq_0^4)\phi _b+\frac16u\phi
  _b^3+hq_0^2\nabla^2\phi_b]\delta \phi\right.\nonumber\\
&+&\left.\frac12(\tau+\frac12u\phi_b^2)\left(\delta\phi\right)^2
+\frac12h\left(\nabla^2\delta\phi+q_0^2\delta\phi\right)^2
\right\}{\rm d}^3{\bf r}\nonumber\\
&+&\int
\left[\sigma\delta\phi+\tau_s\left(2\phi_b\delta\phi+
\delta\phi^2\right)\right]{\rm
d^2{\bf r}_s}
\end{eqnarray}

In the next section we investigate the parallel and perpendicular
lamellar phases denoted as $L_\parallel$ and $L_\perp$,
respectively, and choose the appropriate forms for their bulk
phase $\phi_b$. The free energy, Eq. (\ref{deltaF}), is then
minimized with respect to the correction field $\delta\phi$ and
yields the BCP  profile.

\section{Order parameter profiles}\label{profiles}

The cases of parallel
$L_\parallel$ and perpendicular $L_\perp$ phases are now
considered separately.

\subsection{Perpendicular lamellar phase: $L_\perp$}\label{Lperp}

Up to this point $\delta \phi$ and $\phi_b$ were not specified.
For films below the ODT ($\tau<0$), the perpendicular bulk phase
$L_\perp$ has the bulk periodicity $d_0=2\pi/q_0$. Its order
parameter is given in the single mode approximation (close to the
ODT) by \cite{N-A-SPRL97}
\begin{equation}\label{bulk_per}
\phi_b({\bf r})=\phi_q\cos(q_0 x)
\end{equation}
the amplitude $\phi_q=(-8\tau/u)^{1/2}$ is obtained from a
variational principle of the bulk free energy.

The order parameter for the perpendicular lamellae is
\begin{eqnarray}
\phi_\perp({\bf r})&=&\phi_b({\bf r})+\delta\phi({\bf
r})\nonumber\\
\delta\phi({\bf r})&=&w(y)+g(y)\cos(q_0x)\label{dphi_per}
\end{eqnarray}
where for the correction field $\delta\phi$ we use the single
mode ansatz. If additional modes are included in the bulk order
parameter, Eq. (\ref{bulk_per}), such modes should also be
included in Eq. (\ref{dphi_per}).

For the above choice of $\phi_b$ [Eq.~(\ref{bulk_per})], it is
now possible to perform the $x$ and $z$ integration explicitly,
retaining only the $y$ dependency in Eq.~(\ref{deltaF}). The free
energy per unit area can be written as
\begin{equation}\label{DF_per}
\Delta F_\perp=\Delta F_g+\Delta F_w
\end{equation}
where
\begin{eqnarray}\label{DF_g}
\Delta F_g=\int \left\{-\frac12\tau
g^2+\frac14h\left(g''\right)^2\right\}{\rm
d}y\nonumber\\
+~\tau_s\phi_q(g_++g_-)+\frac12\tau_s(g_-^2+g_+^2)~~~
\end{eqnarray}
and
\begin{eqnarray}\label{DF_w}
\Delta F_w=\int \left\{-\frac12\tau
w^2+\frac12h\left(q_0^2w+w''\right)^2 \right\}{\rm
d}y\nonumber\\
+~\sigma^-w_-+\sigma^+w_++\tau_s(w_-^2+w_+^2)~~~
\end{eqnarray}
where $g_\pm\equiv g(\pm L)$ and $w_\pm\equiv w(\pm L)$.

The amplitude function $g(y)$ results from the surface
modification of the Flory parameter, $\tau_s\lessgtr 0$, and it
vanishes if $\tau_s$ vanishes. This can be seen by noting that if
$\tau_s=0$ then the minimum of the integral in Eq. (\ref{DF_g})
is obtained for $g(y)\equiv 0$ (recalling that $\tau<0$). There
is no coupling between $w(y)$ and $g(y)$, since the free energy
is expanded to second order in $\delta\phi$, and the mixed terms
are of higher order. The function $w(y)$ minimizes $\Delta F_w$
subject to the condition that $\int w(y) {\rm d}y$ is fixed.
Using $\lambda$ as the Lagrange multiplier, it satisfies an
ordinary fourth order differential equation
\begin{equation}\label{gov_perw}
(q_0^4-\frac{\tau}{h})w(y)+2q_0^2w''(y)+w''''(y)
-\frac{\lambda}{h}=0 \end{equation}
Similarly, the equation for $g(y)$ is
\begin{eqnarray}\label{gov_perg}
-\frac{2\tau}{h}g+g''''=0
\end{eqnarray}
A Lagrange multiplier is not needed here because \\$\int
g(y)\cos(q_0x)~{\rm d}^3r=0$. These equations are linear in
$w(y)$ and $g(y)$ since the free energy, Eq.~(\ref{deltaF}), is
expanded to second order around $\phi_b$. The four boundary
conditions for $g(y)$ are
\begin{eqnarray}
2\tau_s\phi_q+2\tau_s g_\pm
\mp hg'''(\pm L)=0\label{bcs_perg12}\\
g''(\pm L)=0\label{bcs_perg34}
\end{eqnarray}
and for $w(y)$
\begin{eqnarray}
\sigma^\pm+2\tau_sw_\pm
\mp q_0^2hw'(\pm L)\mp hw'''(\pm L)=0\label{bcs_perw12}\\
q_0^2 w_\pm + w''(\pm L)=0\label{bcs_perw34}\\
\int_{-L}^{L}w(y){\rm d}y=0\label{bcs_perw5}
\end{eqnarray}
Equation~(\ref{bcs_perw5}) expresses the condition that the total
A/B fraction is conserved, $\int \delta\phi({\bf r}){\rm
d}^3r=0$.

All coefficients in Eqs. (\ref{gov_perw}) and (\ref{gov_perg})
are constants, and therefore the solutions $g(y)$ and $w(y)$ have
the form:
\begin{eqnarray}
w(y)&=& A_we^{-k_w y}+B_we^{k_w y}~+~A_w^*e^{-k_w^* y}
+B_w^*e^{k_w^* y}\nonumber \\ && + ~const.\label{wy_per}\\
g(y)&=& A_ge^{-k_g y}+B_ge^{k_g y}~ +~ A_g^*e^{-k_g^* y}+
B_g^*e^{k_g^* y} \label{gy_per}
\end{eqnarray}
where $O^*$ denotes the complex conjugate of the variable $O$.
The complex amplitudes $A_w$, $B_w$, $A_g$ and $B_g$ and the
constant term in Eq.~(\ref{wy_per}) are determined from the
boundary conditions.

The complex wavevectors $k_w$ and $k_g$ are given by
\begin{eqnarray}
k_w^2&=&-q_0^2+(\tau/h)^{1/2}\\
k_g^2&=&(2\tau/h)^{1/2}
\end{eqnarray}
In the vicinity of the ODT, $\tau\approx 0$, and the real and
imaginary parts of $k_w=k_w'+ik_w''$ are given approximately by
\begin{eqnarray}
k_w '&\approx& \frac{\alpha}{2q_0}\left(N\chi-N\chi_c\right)^{1/2}
\label{real_per}\\
k_w ''&\approx& q_0
\left(1-\alpha^2\frac{N\chi-N\chi_c}{8q_0^4}\right)\label{im_per}
\end{eqnarray}
where $\alpha\equiv 2q_0^2/(1.95\sqrt{3}c)$ follows from Eqs.
(\ref{consts1}) and (\ref{consts2}). The period of modulations
$2\pi/k_w ''$ tends to $2\pi/q_0$, and the decay length of these
modulations $\xi_w=1/k_w'$ diverges as $\xi_w\sim
(N\chi-N\chi_c)^{-1/2}$ in the limit $\tau\to 0$
\cite{mm01,epl01}.

\begin{figure}[h]
\begin{center}
\resizebox{0.45\textwidth}{!}{
 \includegraphics{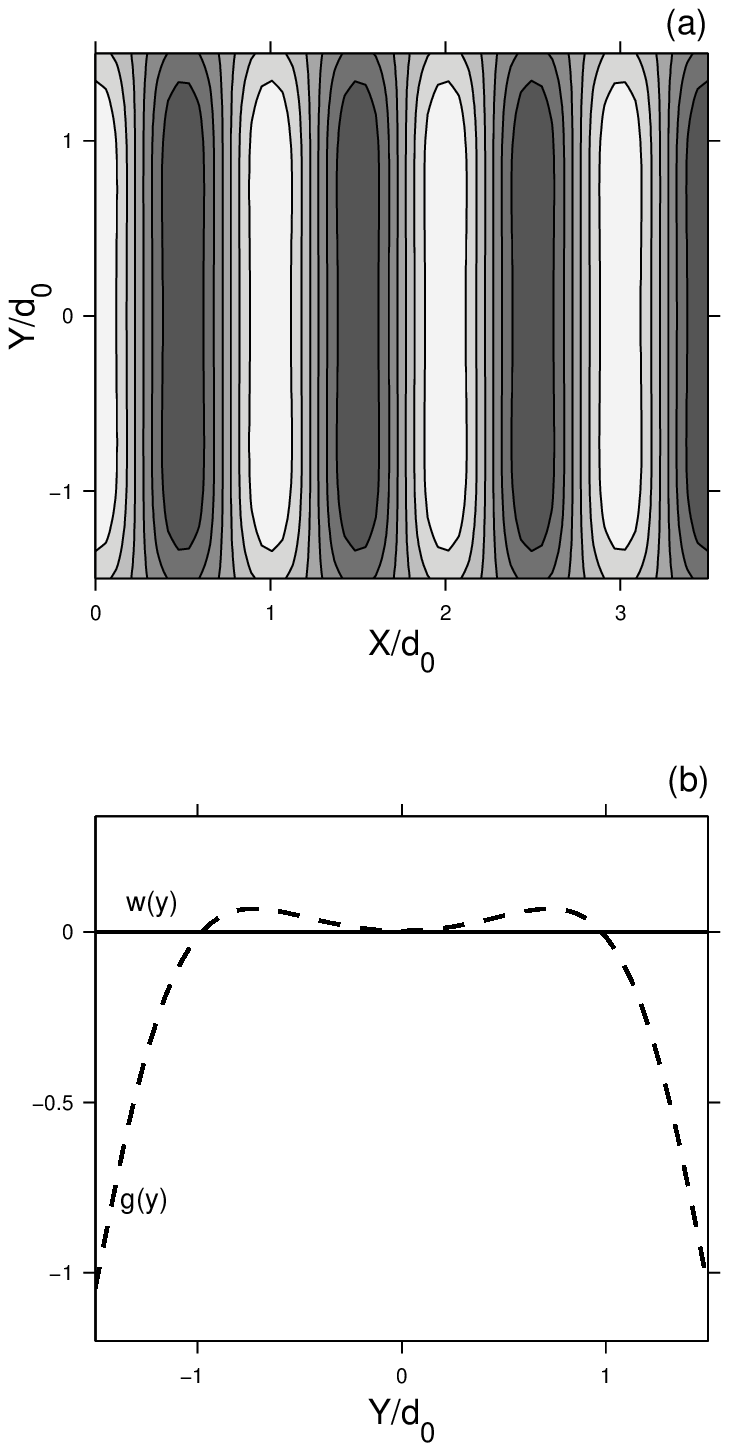}
}
\end{center}
\caption{Contour plot of the perpendicular lamellar phases
between two homogeneous walls. The A monomers are shown in light
shades while the B ones are dark. In (a) the two walls at $y=\pm
L=\pm 1.5d_0$ are neutral, $\sigma^\pm=0$. Part (b) shows the
correction fields $w(y)$ (solid line) and $g(y)$ (dashed line) in
$\delta\phi({\bf r})=w(y)+g(y)\cos(q_0x)$. A surface Flory
parameter which is different from the bulk value, $\tau_s>0$,
causes surface deviations of the lamellar structure from its bulk
shape, even for neutral walls.  The Flory parameter is $\chi
N=10.8$, $\tau_s=0.1hq_0^3$. In this and subsequent figures we
use $u/\rho=c=1$ and $N=1000$.}
\label{Fig. 1}
\end{figure}

A contour plot of the order parameter $\phi(x,y)=\\
\phi_b(x,y)~+~\delta\phi(x,y)$ is shown in Fig.~1(a), for
inter-plate separation $2L=3d_0$. The two walls at $y=\pm L$ are
neutral, $\sigma^\pm =0$, but the surface Flory parameter
deviates from its bulk value, $\tau_s>0$. Note that the
interfacial width broadens close to the wall, but the A/B
inter-material dividing surface (IMDS) (defined as the surface
where $\phi(x,y)=0$) is perpendicular to the walls. This result
is similar to the one obtained in Ref. \cite{G-M-B99} (their
Fig.~3), using different methods. In Fig.~1(b) we show the
response fields $g(y)$ and $w(y)$ in
$\delta\phi=w(y)+g(y)\cos(q_0x)$. It is advantageous for the
lamellae to reduce their amplitude close to the wall, hence, in
our convention, a positive $\tau_s>0$ induces a negative
$g(y=0)$. The amplitude of sinusoidal modulations in
$\phi(x,y)=w(y)+(\phi_q+g(y))\cos(q_0x)$ is therefore diminished
from it unperturbed value $\phi_q$. In the absence of surface
fields, $\sigma^\pm$, the $w$ part of $\delta\phi$ vanishes,
$w(y)=0$.

\begin{figure}[h]
\begin{center}
\resizebox{0.45\textwidth}{!}{
 \includegraphics{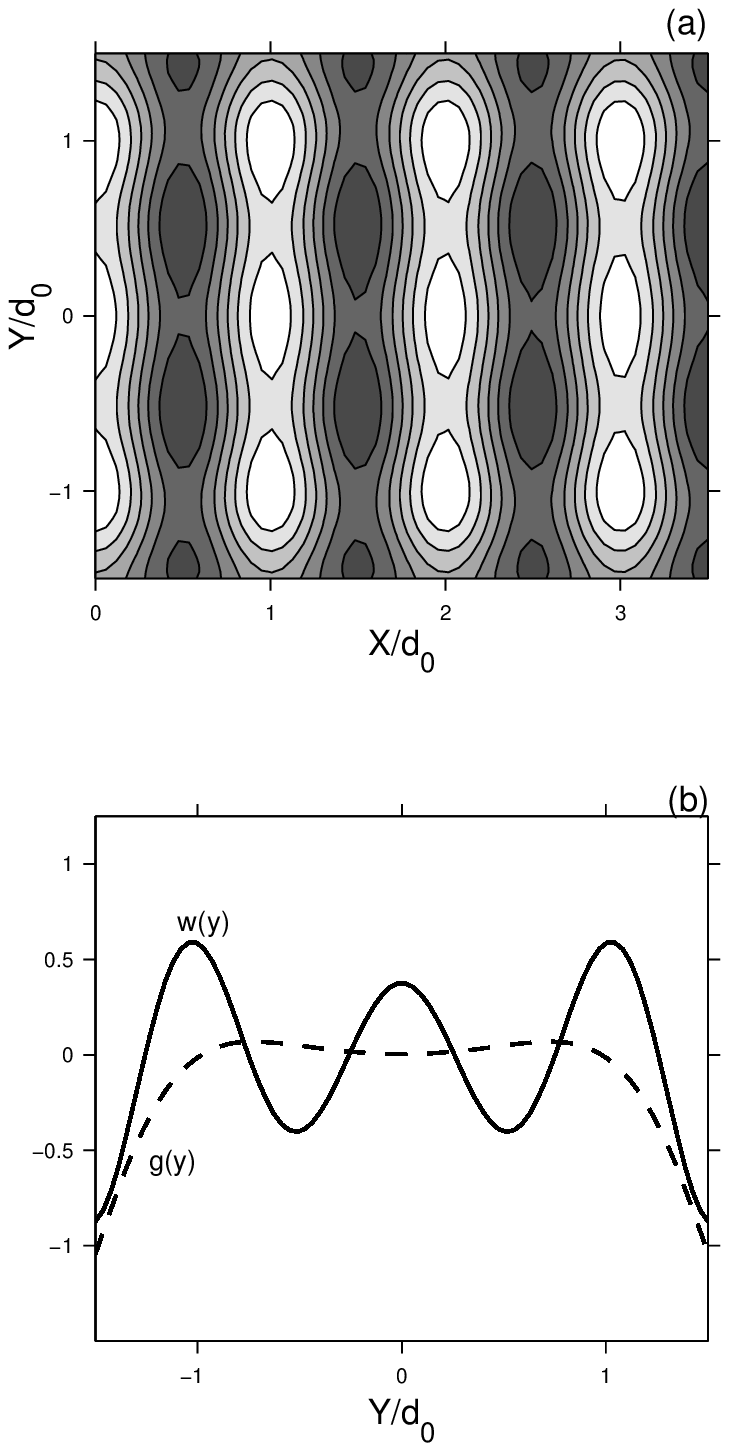}
}
\end{center}
\caption{Same as in Fig.~1, but here the two walls favor the B
monomers, $\sigma^\pm=0.2hq_0^3\phi_q>0$. Monomers are rearranged
near the walls and the A/B inter-material dividing surface (IMDS)
is curved (see also Fig.~3). The preferential walls induce
parallel ordering, as $w(y)\neq 0$ in (b). The length scale of
modulation in (a) is determined by the functions $w(y)$ and
$g(y)$ in (b) [Eqs. (\ref{wy_per}) and (\ref{gy_per})].}
\label{Fig. 2}
\end{figure}
Figure~2 is similar to Fig.~1, but the symmetric walls
($\sigma^+=\sigma^-$) are chosen here to favor the B monomers (in
dark), which partially wet them. As a result, the A/B IMDS bends
and intersects with the walls at an angle which is different than
$90^\circ$. The preferred adsorption is also seen in Fig.~2(b),
where $w(y)$ is negative at the walls, $w_\pm<0$.

The copolymer contour lines are defined by the relation
$\phi({\bf r})=\phi_b({\bf r})+\delta\phi({\bf r})=c$, where $c$
is a constant. Clearly, the inter-material dividing surface
(IMDS) is just the special case with $c=0$. For bulk lamellar
phase the IMDS are just parallel planar surfaces (lines in two
dimensions), but for lamellae confined in thin films the shape of
these lines is more complicated. Figure~3(a) shows the IMDS lines
for symmetric walls, both favoring the B monomer.  As expected,
the contact area of the B domains with the wall is increased, and
the IMDS lines are curved appropriately.  A different behavior is
seen in Fig.~3(b) (asymmetric walls) where the curving of the
IMDS lines is opposite at the two surfaces because of the
opposite wall interaction. The deviation from a perfect lamellar
shape is seen as the IMDS undulates.

In general, contour lines do not run perpendicular to the wall
but rather form an angle different than 90$^\circ$ with the
surface.
%
\begin{figure}[h]
\begin{center}
\resizebox{0.4\textwidth}{!}{
 \includegraphics{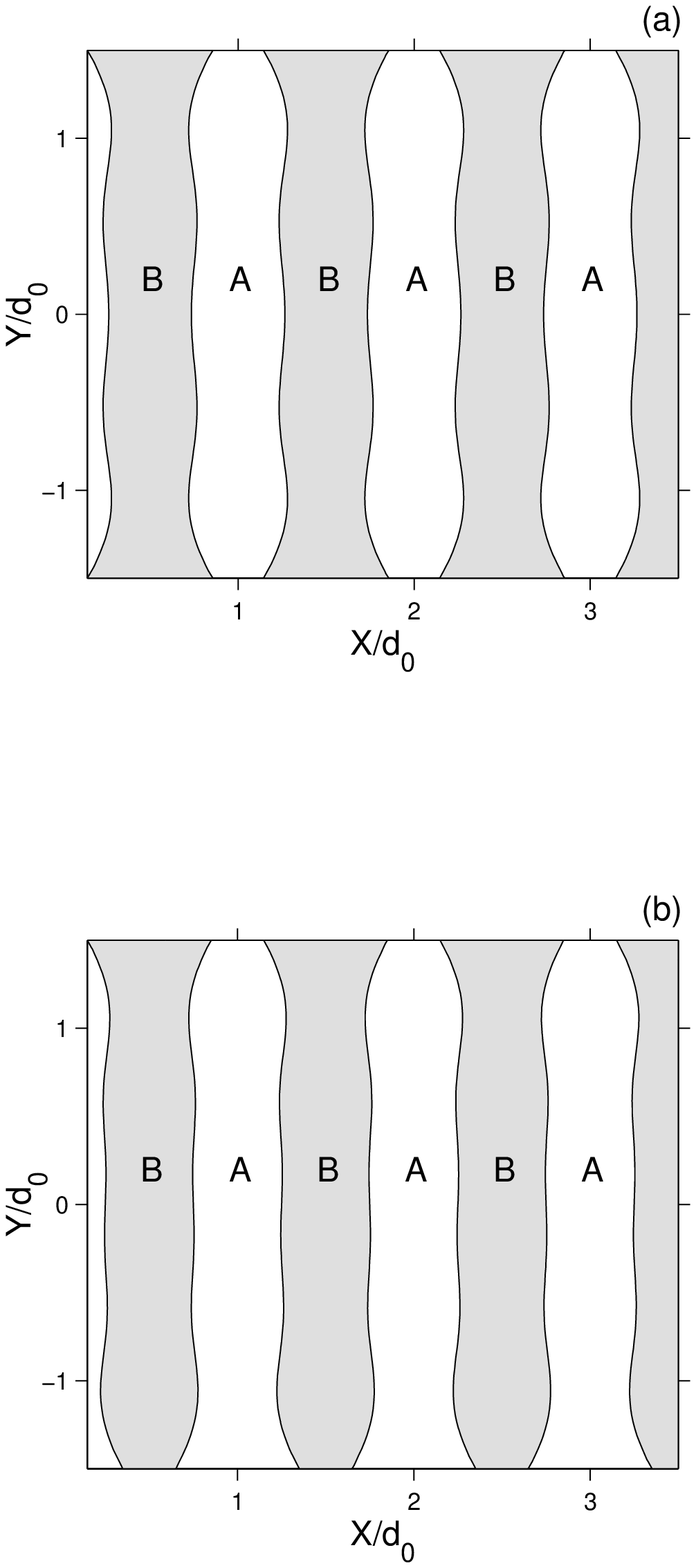}
}
\end{center}
\caption{Parts (a) and (b) are plots of the IMDS (defined by
$\phi({\bf r})=0$) of confined perpendicular $L_\perp$ lamellae.
In (a), the two walls favor the B monomers,
$\sigma^\pm=0.2hq_0^3\phi_q>0$, and the B domains are larger than
the A domains at the walls. In (b)
$\sigma^-=-\sigma^+=0.2hq_0^3\phi_q$, and the A domains have
large size at $y=-L$, while the B domains are larger at $y=L$.
The Flory parameter is $N\chi=11$ and $\tau_s=0.1hq_0^3$.}
\label{Fig. 3}
\end{figure}
On contour lines having $\phi=const.$, $x$ and $y$ are related by
\begin{equation}\label{imds1}
\cos q_0x=\frac{c-w(y)}{\phi_q+g(y)}
\end{equation}
Figure 4 is a schematic presentation of the IMDS. The dotted
vertical line shows the unperturbed location of the A/B IMDS. At
the $y=-L$ wall, the deviation ${\rm \Delta} x$ of the IMDS from this
line (see Fig.~4) is
\begin{equation}
\frac{{\rm \Delta} x}{d_0}=\frac{1}{2\pi}
\arccos(\frac{-w_-}{\phi_q+g_-})-\frac14
\end{equation}
The
departure from the flat interface can be quite large, for
example, in Fig.~3(a) it is ${\rm \Delta} x/d_0\approx 0.1$.
\begin{figure}[h]
\begin{center}
\resizebox{0.4\textwidth}{!}{
 \includegraphics{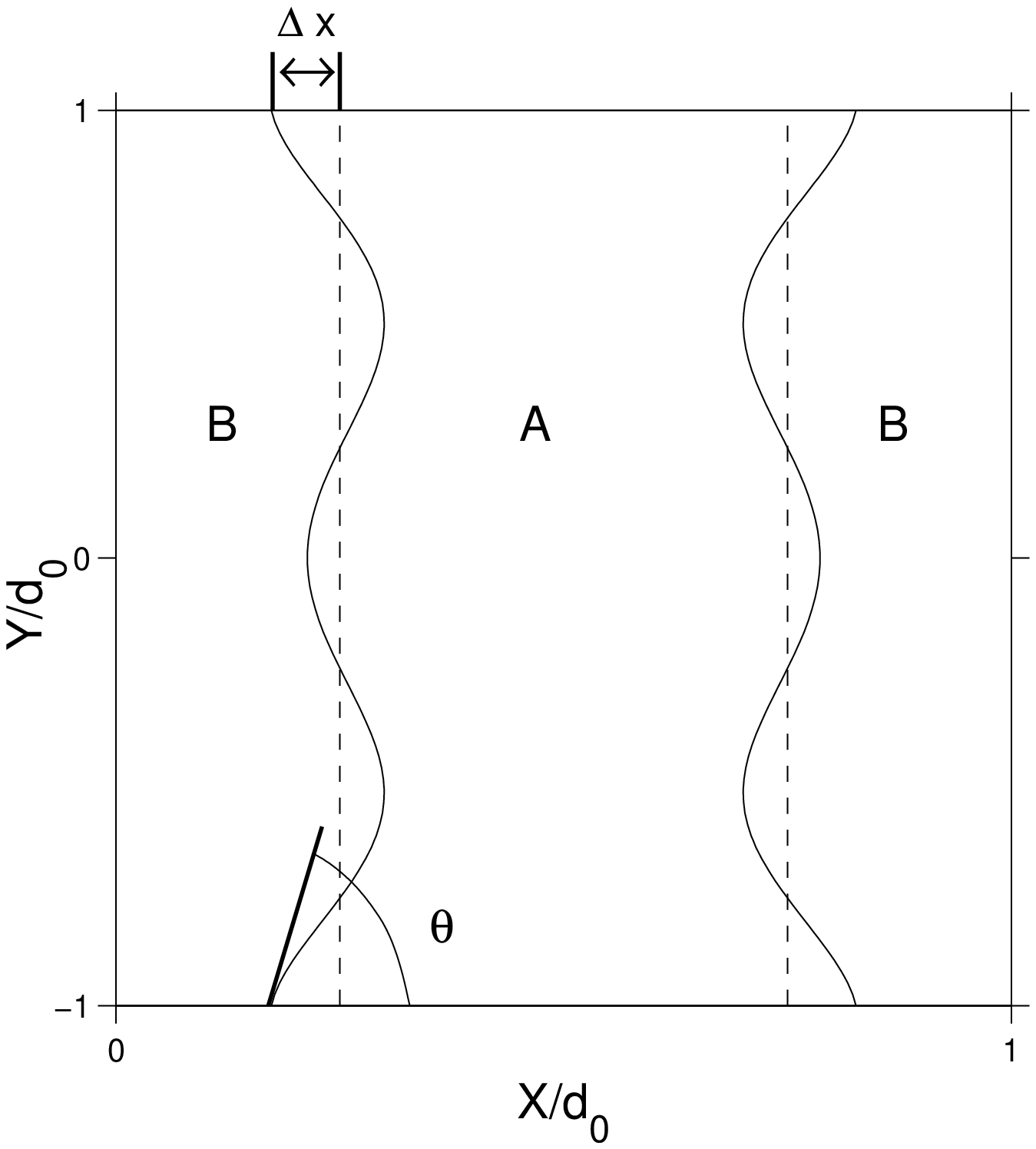}
}
\end{center}
\caption{Schematic drawing of the IMDS lines. The confining walls
are at $y=\pm L=\pm d_0$. The dotted line is the location of the
unperturbed IMDS. The lateral deviation from this line at the
walls is ${\rm \Delta} x$. The angle between the tangent to the
IMDS and the $x$-axis is $\theta$.}
\label{Fig. 4}
\end{figure}
We define $\theta$ as the angle at which the IMDS line $y(x)$
joins the wall at $y=-L$. For neutral walls, ${\rm \Delta }x=0$
and $\theta=90^\circ$. From Eq. (\ref{imds1}) it follows that
\begin{equation}
\tan \theta=\frac{{\rm d}y}{{\rm d}x}
=\frac{q_0\sin (q_0\Delta
x )(\phi_q+g_-)^2}{w'(-L)(\phi_q+g_-) +w_-g'(-L)}
\end{equation}
Using the same parameters as in Fig.~3(a), we find that $\theta\approx
80^{\circ}$.

As the ODT is approached from below ($\tau<0$), the lamellae can
be deformed more easily. The energetic cost of lamellae bending
and compression is reduced, and the IMDS departs appreciably from
its flat shape. The effect of temperature is clearly seen in
Fig.~5, where in (a) the IMDS is plotted for Flory parameter
$\chi N=12$, while in (b) the temperature is higher and closer to
the ODT, $\chi N=11$, and the contour lines show stronger
undulations. Close to the ODT, the length scale associated with
the undulation periodicity is $2\pi/k_w''\approx d_0$ [see Eq.
(\ref{im_per})], but it may get much smaller as the temperature
is reduced, $\chi\gg\chi_c$. The second length scale in the
system, $2\pi/k_w'$, characterizes the decay of modulations, and
it diverges at the ODT.

\begin{figure}[h]
\begin{center}
\resizebox{0.4\textwidth}{!}{
 \includegraphics{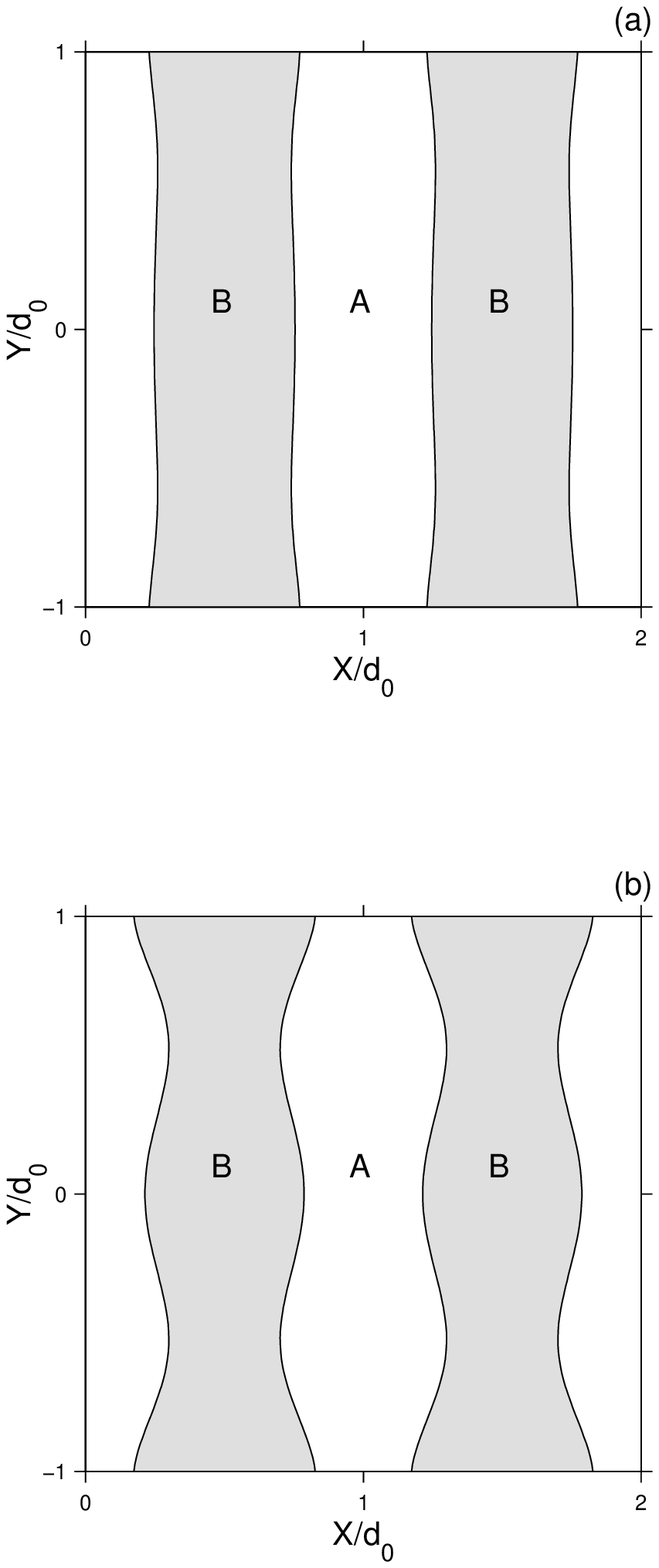}
}
\end{center}
\caption{Temperature dependence of the shape of the IMDS. In (a)
the Flory parameter is $\chi N=12$ (relatively strong
segregation), and the IMDS are nearly flat. As the temperature is
raised and approaches the ODT, $\chi N=11$ in (b), the lamellae
can easily deform in accordance with the surface fields
$\sigma^\pm$. The shape of decaying undulations is given by Eq.
(\ref{imds1}) with $c=0$. The parameters chosen are
$\sigma^\pm=0.5hq_0^3$ and $\tau_s=0$.}
\label{Fig. 5}
\end{figure}

\subsection{Parallel lamellar phase: $L_\parallel$}

The alternative case of lamellar order occurs when the lamellae
are parallel to the walls, and the A/B profiles depend only on
the distance from the walls, $\phi({\bf r})=\phi(y)$. In the
strong stretching approximation \cite{W-RMM94,turnerPRL92}, the
lamellae are allowed to stretch or compress in order to vary
their natural periodicity $d_0$ according to the constraint
inter-plate separation $2L$. The system can have $n$ or $n\pm
1/2$ lamellae between the walls, where $n$ is the closest integer
to $2L/d_0$. This strong stretching calculation motivates our
zeroth order approximation to the $L_\parallel$ phase,
\begin{equation}
\phi_b({\bf r})=\pm\phi_q\cos[q(y+L)]
\label{bulk_par}
\end{equation}
Using $n$ from above, the wavenumber is $q=n\pi/L$ or
$(n+1/2)\pi/L$. The lamellae are stretched if $q<q_0$ and
compressed if $q>q_0$. The $\pm$ sign of the profile is
determined by the wall interactions.

The bulk approximation for the profile, Eq. (\ref{bulk_par}),
serves as a starting point. However, the correction field,
$\delta\phi$, has an important contribution in the weak
segregation. The order parameter for the parallel phase
$L_\parallel$ is
\begin{eqnarray}
\phi_\parallel({\bf r})&=&\phi_b({\bf r})+\delta\phi({\bf r})\nonumber\\
\delta\phi({\bf r})&=&w(y)\label{dphi_par}
\end{eqnarray}
The free energy (per unit area) has $y$-dependent terms only.
Expanded to second order in $w(y)$, it can be written as:
\begin{eqnarray}\label{DF_par}
\Delta F_{\parallel}&=&\int\left[ \frac12(\tau
+\frac12u\phi_b^2)
w^2
+~\frac12h\left(q_0^2w+w''\right)^2\right]{\rm
d}y\nonumber\\
&+&~\sigma^-w_-+2\tau_s\phi_b(-L)w_-+\tau_sw_-^2\nonumber\\
&+&~\sigma^+w_++2\tau_s\phi_b(L)w_++\tau_sw_+^2
\end{eqnarray}
Similar to the treatment of the perpendicular phase in Sec. \ref{Lperp},
this free energy is minimized to yield a linear
differential equation, but with $y$-dependent coefficients:
\begin{eqnarray}\label{gov_par}
w^{\prime\prime\prime\prime}(y)+2q_0^2w^{\prime\prime}(y)&&
\nonumber\\
+\left[q_0^4-\frac{\tau}{h}-\frac{\tau}{h}\cos(2qy)\right]w(y)-
\frac{\lambda}{h}&=&0
\end{eqnarray}
The conditions imposed on $w(y)$ are:
\begin{eqnarray}
\sigma^\pm+2\tau_s\phi_b(\pm
L)+2\tau_sw_\pm&&\nonumber\\
\mp~ q_0^2hw'(\pm L)\mp hw'''(\pm L)&=&0\\
\nonumber\\
q_0^2w(\pm L)+w''(\pm L)&=&0\\
\int_{-L}^L w(y){\rm d}y&=&0
\end{eqnarray}
where as before $\lambda$ is the Lagrange multiplier and the last
equation expresses the conservation of the relative A/B
concentration in the film. The homogeneous solution of
Eq.~(\ref{gov_par}) has the Bloch (Floquet) form
\begin{equation}\label{bloch}
w(y)=e^{-ky}\sum_na_ne^{2inqy}+e^{-k^*y}\sum_na_n^*e^{-2inqy}
\end{equation}
A recurrence relation between the coefficients $\{a_n\}$ is
obtained by substituting Eq. (\ref{bloch}) in
Eq.~(\ref{gov_par}). However, the recurrence relation converges
only for specific values (eigenvalues) of $k$. If $k$ is a valid
eigenvalue, then so are $k^*$, $-k$ and $-k^*$. These four
eigenvalues correspond to the four independent solutions of the
fourth order differential equation, Eq.~(\ref{gov_par}).

A useful approximation to the free energy Eq.~(\ref{DF_par}) is
obtained by replacing $\phi_b^2$ by its average, smeared value
$\langle\phi_b^2\rangle$. This is equivalent to replace the
potential term \\ $-(\tau/h)\cos2qy$ ~ in Eq.~(\ref{gov_par}) by
its zero average. The governing equation for the correction field
$w(y)$ is then given by a linear differential equation with
constant coefficients
\begin{equation}\label{gov_app}
w''''(y)+2q_0^2w''(y)+(q_0^4-\tau/h)w(y)-\lambda/h=0
\end{equation}
Under this approximation, the form of $w(y)$ in the parallel
$L_\parallel$ phase is the same as it is in the perpendicular
$L_\perp$ phase, Eq. (\ref{wy_per}), only the boundary conditions
are different.

\begin{figure}[h]
\begin{center}
\resizebox{0.45\textwidth}{!}{
 \includegraphics{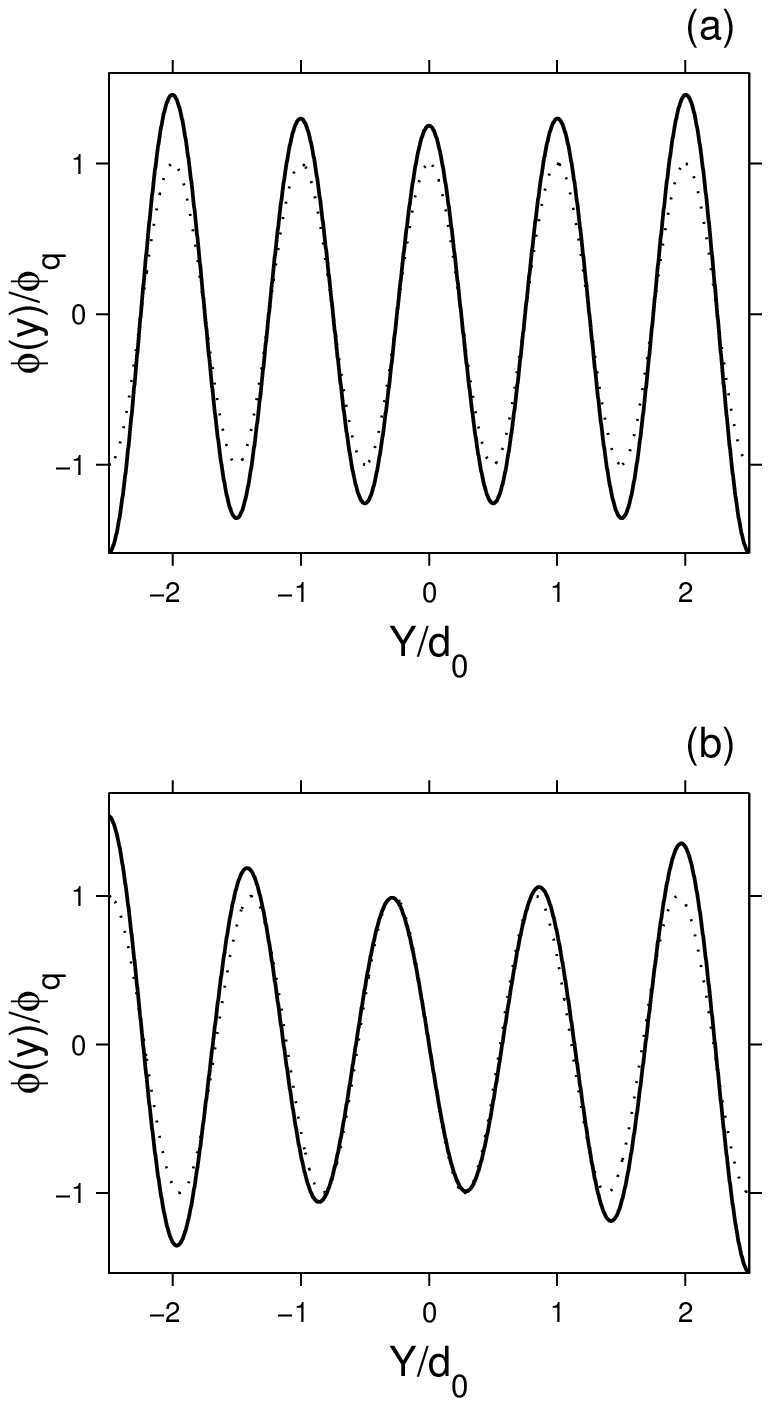}
}
\end{center}
\caption{Concentration profiles for the confined parallel
$L_\parallel$ phase. Dotted line [$\phi=\phi_b(y)$] and solid
line [$\phi=\phi_b(y)+\delta\phi(y)$] are normalized by $\phi_q$.
In (a) the two walls favor the B monomers ($\phi<0$),
$\sigma^\pm=0.5hq_0^3\phi_q>0$ and the film is symmetric, while
in (b) the film is asymmetric,
$\sigma^-=-\sigma^+=-0.5hq_0^3\phi_q$, and the A monomers are
adsorbed at the $y=-L$ wall. The bulk Flory parameter is $\chi
N=10.6$ and its surface modification is $\tau_s=0.125hq_0^3$. }
\label{Fig. 6}
\end{figure}

Order parameter profiles are presented in Fig.~6. The dotted line
is $\phi_b$ as obtained by the bulk approximation,
Eq.~(\ref{bulk_par}), and the solid line is the full profile,
$\phi=\phi_b(y)+\delta\phi(y)$. In Fig.~6(a) the interfacial
interactions are the same on both walls, $\sigma^+=\sigma^-$,
inducing a symmetric lamellar ordering. The difference between
the two curves is the correction field $\delta\phi(y)$, favoring
adsorption of the B monomers ($\phi<0$) at the two walls. In 6(b)
the film is asymmetric with $\sigma^+=-\sigma^-$, and adsorption
of the A monomers at the $y=-L$ wall is enhanced.

\section{Free energy and stability diagram}\label{energy}

Once the order parameter profiles for the parallel and
perpendicular lamellar phases are calculated, the corresponding
free energies can be evaluated by substituting the order
parameter profiles in Eqs. (\ref{DF_par}) and (\ref{DF_per}),
respectively. The {\it reference} free energy $F^0[\phi_b]$ is
calculated by the bulk approximation. For the parallel lamellae
it is given by substituting the profile, Eq. (\ref{bulk_par}),
directly into Eq. (\ref{Fb}),
\begin{eqnarray}\label{f0_par}
F_\parallel^0[\phi_b]=\left[\frac14
\tau\phi_q^2+\frac14h(q_0^2-q^2)^2\phi_q^2
+\frac{u}{64}\phi_q^4\right]2L\nonumber\\
\pm~\sigma^-\phi_q\pm\sigma^+\phi_q+2\tau_s\phi_q^2
\end{eqnarray} %
where the $\pm \sigma$ terms result from the choice of the bulk
order parameter $\phi_b$. For the perpendicular phase $L_\perp$,
substituting the profile , Eq. (\ref{bulk_per}), results in:
\begin{eqnarray}\label{f0_per}
F_\perp^0[\phi_b] &=&\left[\frac14\tau\phi_q^2
+\frac{u}{64}\phi_q^4\right]2L+\tau_s\phi_q^2\nonumber\\
&=&-\frac{2L\tau^2}{u}+\tau_s\left(\frac{-8\tau}{u}\right)
\end{eqnarray}
\begin{figure}[h]
\begin{center}
\resizebox{0.45\textwidth}{!}{
 \includegraphics{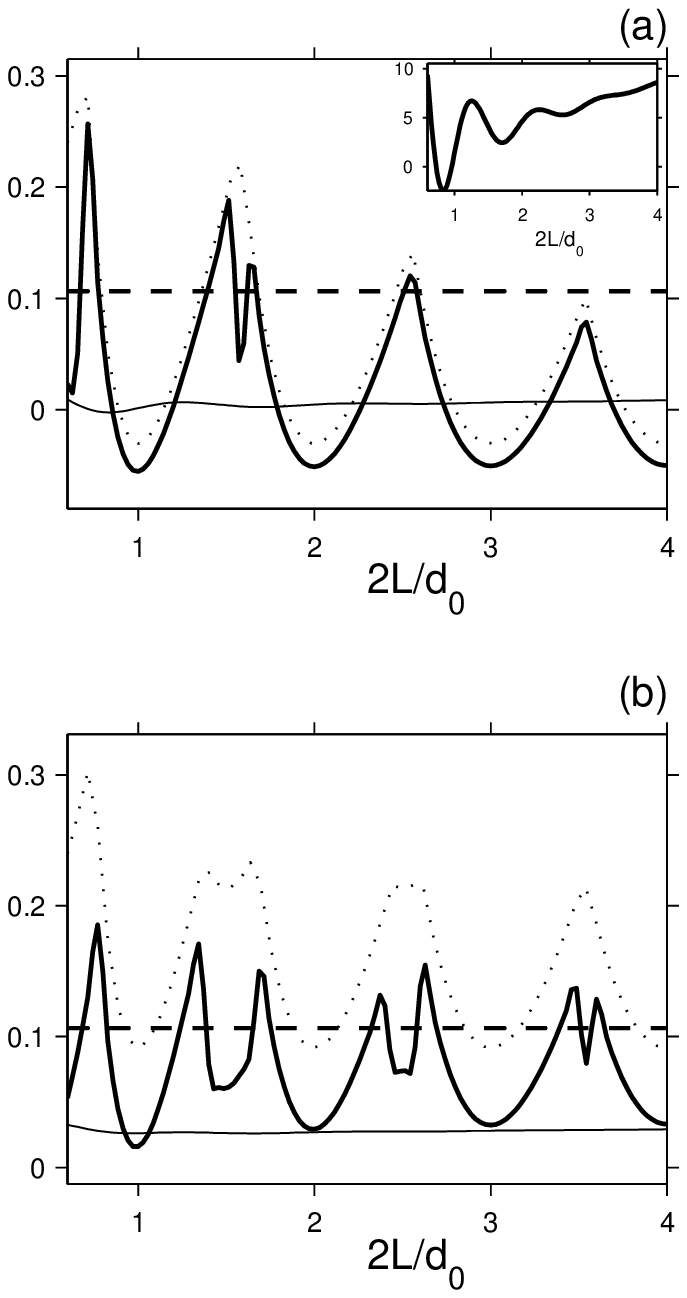}
}
\end{center}
\caption{Film free energy per unit area as a function of
inter-plate separation $2L$. Shown are the bulk approximation to
the free energy of the $L_\perp$ phase (horizontal dashed line),
bulk approximation of $L_\parallel$ (dotted line), full free
energy of the $L_\perp$ (solid line) and of the $L_\parallel$
phase (thick solid line). The film is taken to be symmetric. In
(a) $\sigma^\pm =0.4hq_0^3\phi_q$, while in (b) the surface
interactions are smaller, $\sigma^\pm=0.2hq_0^3\phi_q$. Free
energies are measured with respect to the free energy of the bulk
lamellar phase. Inset in (a) is an enlargement of the $L_\perp$
free energy by a factor of $10^3$, showing a deep minimum for
$2L\lesssim d_0$. The bulk Flory parameter is $\chi N=11$ and its
surface deviation is $\tau_s=0.35hq_0^3$.}
\label{Fig. 7}
\end{figure}

As a function of inter-plate separation $2L$, the total free
energy $F[\phi_b+\delta\phi]$ has oscillations, as depicted in
Fig.~7 for symmetric film, $\sigma^-=\sigma^+$. The free energies
of the perpendicular and parallel lamellar phases (solid line and
thick solid line, Eqs.~(\ref{DF_per}) and (\ref{DF_par}), are
lower than the bulk ones $F_\parallel^0[\phi_b]$ and
$F_\perp^0[\phi_b]$ (dotted and dashed lines). In Fig.~7(a) the
wall interactions are $\sigma^\pm =0.4hq_0^3\phi_q$, and the free
energy of the $L_\parallel$ phase is slightly reduced from the
bulk approximation value. Additional minimum develops at
$2L\approx 1.5d_0$. The $L_\perp$ free energy has a marked
minimum for $2L\lesssim d_0$ \cite{P-BMM97}, see inset. In
Fig.~7(b) the interfacial interactions are smaller,
$\sigma^\pm=0.2hq_0^3\phi_q$, and in this case the $L_\parallel$
free energy is notably lowered from the bulk approximation
calculation. However, the difference between the two curves tends
to zero as $2L/d_0\to \infty$, because the surface induced
modulations have finite range. For both choices of $\sigma$, the
$L_\perp$ free energy is significantly lowered from its bulk
approximation value. Note that the bulk approximation curves are
similar to the curves obtained in the strong stretching
approximation.
\begin{figure}[h] \begin{center}
\resizebox{0.45\textwidth}{!}{
 \includegraphics{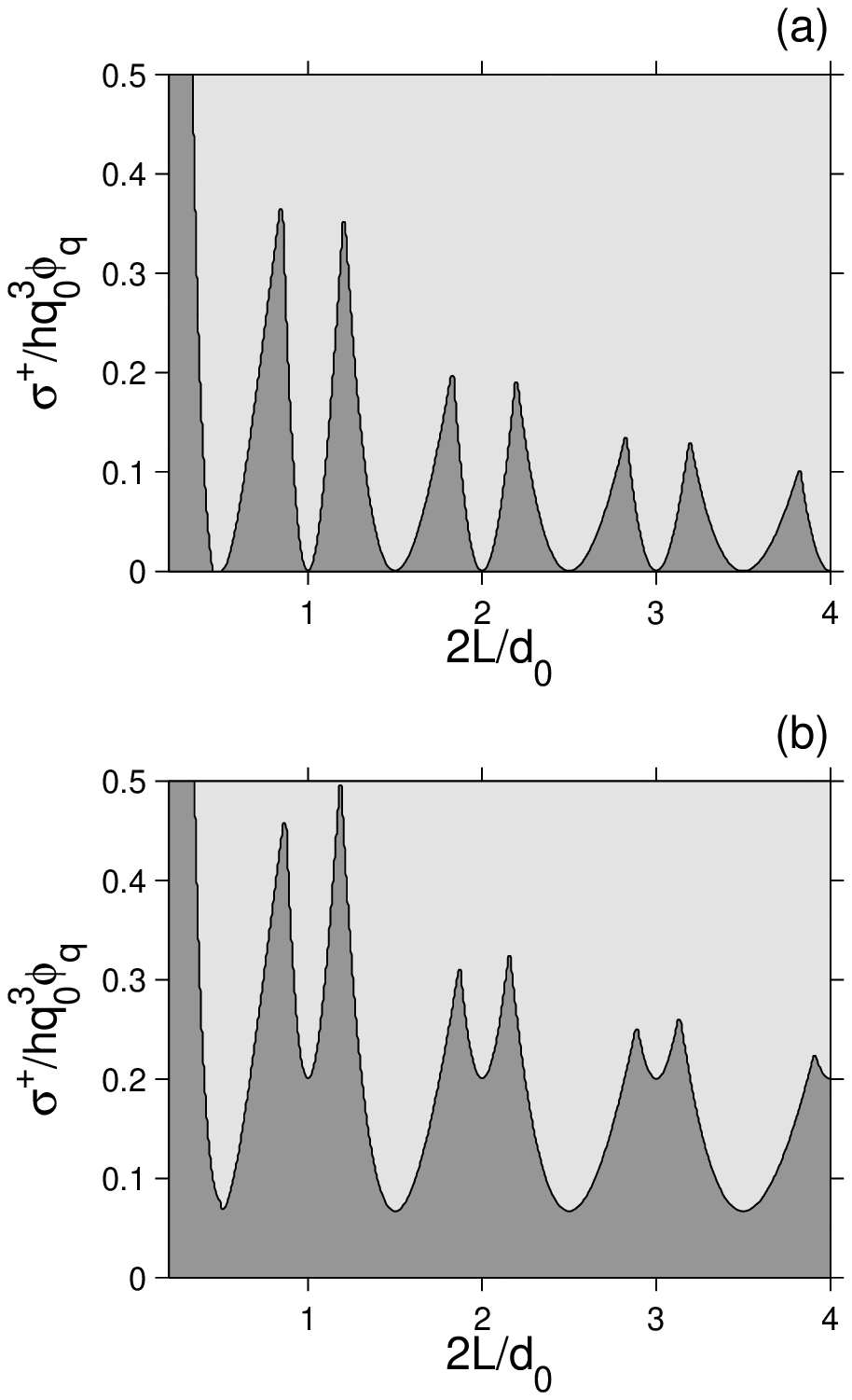}
}
\end{center}
\caption{The stability of $L_\parallel$ (in light) vs. $L_\perp$
lamellae (in dark), as a function of wall separation $2L$ and
interfacial strength $\sigma^+$. The free energies are taken from
Eqs. (\ref{f0_par}) and (\ref{f0_per}), respectively. In (a) the
surface Flory parameter is the bulk one, $\tau_s=0$, while in (b)
$\tau_s=0.1hq_0^3>0$. The $L_\parallel$ phase is pushed upward in
the stability diagram in (b), removing the degeneracy between
$L_\perp$ and $L_\parallel$ that occurs for neutral walls
($\sigma^\pm=0$) when $\tau_s=0$. The calculation is done by the
bulk approximation, $\phi({\bf r})=\phi_b({\bf r})$. The ratio
$\sigma^+/\sigma^-=-2$ is kept constant and the Flory parameter
is $\chi N=11$.}
\label{Fig. 8}
\end{figure}

Restricting ourselves to $L_\parallel$ and $L_\perp$ lamellar
phases, the stability diagram is constructed as a function of two
system parameters: the inter-wall separation $2L$ and the surface
preference $\sigma^\pm$. In the first stage, we ignore the
correction presented above, and use $F_\parallel^0[\phi_b]$ and
$F_\perp^0[\phi_b]$ as given by the bulk approximation
calculation. The stability diagram in Fig.~8 is calculated for
walls having a fixed ratio of surface interaction
$\sigma^+=-2\sigma^-$. Parallel lamellae at $2L/d_0=n$, for
integer $n$, have symmetric ordering, while antisymmetric
ordering occurs for $2L/d_0=n+1/2$. The difference in the diagram
is caused by the choice of $\sigma$'s. In 8(a) the surface Flory
parameter is the same as the bulk one, $\tau_s=0$. For neutral
walls, $\sigma^\pm=0$, the perpendicular lamellae (in dark) are
stable. A degeneracy between $L_\perp$ and $L_\parallel$ phases
occurs for $2L/d_0=n$ or $n+\frac{1}{2}$ ($n=0$, $1$, $2$ ...),
where the parallel lamellae are not frustrated and $q=q_0$. The
parallel lamellae (in light color) are preferred if the surface
interaction is strong enough to overcome the lamellar stretching
or compression. The use of a weak segregation bulk approximation
agrees with previously obtained stability diagrams in
intermediate and strong segregations \cite{matsenJCP97,G-M-B99}.

\begin{figure}[h]
\begin{center}
\resizebox{0.45\textwidth}{!}{
 \includegraphics{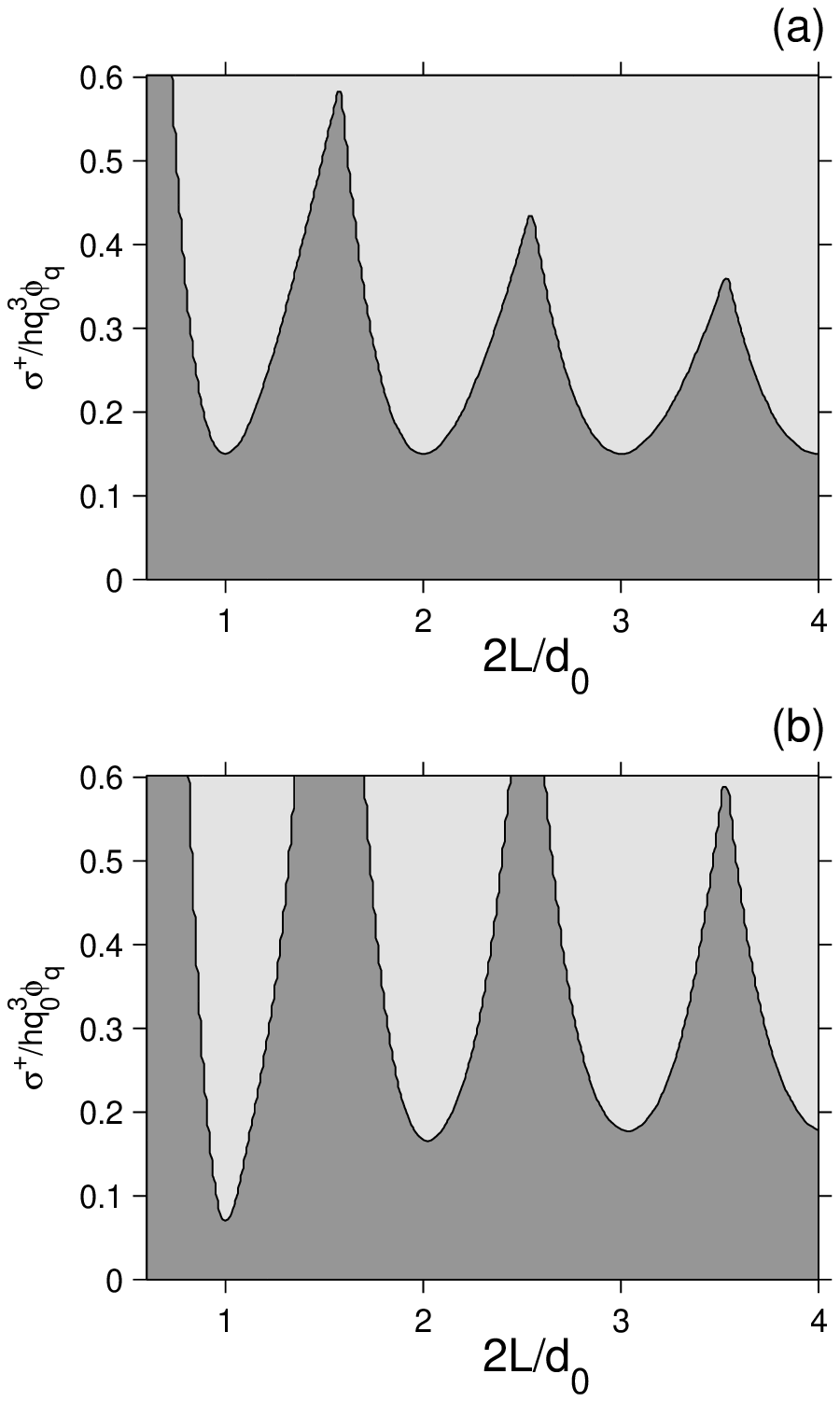}
}
\end{center}
\caption{Stability of $L_\parallel$ (in light) vs. $L_\perp$
lamellae (in dark), comparing in (a) the bulk approximation [free
energy equations (\ref{f0_par}) and (\ref{f0_per})],  with the
full free energy  in (b) [calculated from Eq. (\ref{Fb}) with
Eqs. (\ref{dphi_par}) and (\ref{dphi_per}), respectively]. Note
that the large $\sigma^+$ behavior lies outside the range of
validity of our linear model. In both parts (a) and (b)
$\tau_s=0.3hq_0^3$, the Flory parameter is $\chi N=10.8$ and the
film is symmetric, $\sigma^-=\sigma^+$.}
\label{Fig. 9}
\end{figure}

In Fig.~8(b) we present the bulk approximation, but now the
surface Flory parameter is changed, $\tau_s=0.1hq_0^3>0$. The
$L_\parallel$ phase is pushed upward and the diagram is
different. Symmetric phases [$2L\approx nd_0$] are pushed more
than the asymmetric ones [$2L\approx (n+\frac12)d_0$] because of
our choice of surface fields $\sigma$'s. In the bulk
approximation, the free energy of unfrustrated parallel lamellae
[$q=q_0$ in Eq. (\ref{f0_par})] is higher than that of the
perpendicular lamellae [Eq. (\ref{f0_per})] if the walls are
neutral. As a result, the $L_\perp$ morphology is favored for all
separations $2L$, and the degeneracy is removed
\cite{matsenJCP97,G-M-B99}. Clearly, a surface segregation
temperature different than the bulk one ($\tau_s\neq 0$) can
account for the experimental lack of this degeneracy
\cite{H-RMM98}. According to the same reasoning, if $\tau_s<0$
then the parallel phases are preferred on the expense of the
$L_\perp$ phases, and, in particular, for wall separations
$2L\approx nd_0$. Note that the last sentence agrees with the
different dependence on $\tau_s$ seen in Eqs. (\ref{f0_par}) and
(\ref{f0_per}). For parallel lamellae, the surface term is
$2\tau_s\phi_q^2$, while for perpendicular lamellae it is only
$\tau_s\phi_q^2$.

Figure~9 shows the stability diagram, where in (a) we use the
bulk approximation for symmetric $\sigma^-=\sigma^+$ film, and in
(b) we used the full, and correct, order parameter
$\phi=\phi_b+\delta\phi$. The parallel ordering is then always
symmetric. A general feature of this diagram is that the
$L_\perp$ phase is more stable relative to the $L_\parallel$ for
larger surface fields for $2L>d_0$. The figure also demonstrates
the qualitative agreement with the bulk approximation.

\section{Conclusions}

We have used a Ginzburg-Landau free energy to study analytically
the thin-film ordering of diblock copolymers (BCP) in the weak
segregation regime. The two homogeneous confining walls are
assumed to have short-range interactions with the BCP blocks. The
free energy is expanded to second order around the appropriately
chosen bulk phase, and the correction field $\delta\phi$ is
obtained. The use of such free energy formulation is advantageous
because it offers simple analytical results and complements
numerical studies. However, our mean-field approach is limited to
a region of temperatures in the vicinity of the ODT, but not too
close to it, where critical fluctuations are known to be
important \cite{Semenov85}. Very close to the ODT, the response
field $\delta\phi$ diverges. However, if the surface Flory
parameter is different from the bulk one, $\tau_s>0$, the surface
has a lower ordering temperature than the bulk, and this
divergence is absent \cite{remark}.

For confined parallel $L_\parallel$ and perpendicular $L_\perp$
phases, the correction field $\delta\phi$ adds an enrichment
layer of the preferred component, with thickness $\xi\sim 1/k '$
diverging at the ODT. This thickness is obtained as a special
case for patterned walls (inhomogeneous $\sigma$) studied by us
before. Effects of finite chain length, however, preclude the
divergence of this thickness. In the $L_\perp$ phase, an increase
of the surface fields $\sigma^\pm$ increases the correction field
$\delta\phi$, and induces a parallel lamellar ordering until,
eventually, there is no clear distinction between $L_\parallel$
and $L_\perp$.

In general, the IMDS lines are bent and deviate from their flat
shape in bulk lamellar system. Previous works used a
phenomenological model valid in the strong segregation regime,
and obtained a linear equation for the deviation of the IMDS. The
resulting order parameter expressions for the confined phases are
crude, when compared to Monte-Carlo simulations
\cite{wang00,pwMM,pwPRL}. In the weak segregation presented here,
the order parameter itself is linearized. Using the expressions
given above for $\phi_b({\bf r})$ and $\delta\phi({\bf r})$, one
can deduce the shape of an arbitrary equi-$\phi$ line given by
$\phi({\bf r})=c$. We give expressions for the angle of the IMDS
with the confining walls, and the deviation of the IMDS from its
flat shape. This deviation, characterized by decaying
oscillations, can be quite large and can even reach $20\%-30\%$
of the lamellar width $d_0$. We note that in an experimental
setup whose target is to produce perpendicular lamellae, system
parameters should be tuned in order to keep the lamellae as flat
and parallel as possible.

The free energy as a function of wall separation $2L$ is
different from the bulk approximation. The free energy of the
$L_\perp$ phase is lower than the one obtained the bulk
approximation, as is seen in Fig.~7(a). The curve has decaying
oscillations and tends to a constant when $2L\gg d_0$. The
correction to the $L_\parallel$ free energy has similar
undulatory character and under different conditions its effect
can be large, as in Fig.~7(b). The pressure, $-\partial
F/\partial y$, is different than what is expected from the bulk
approximation, since additional maxima and minima are present in
the free energy. Our bulk approximation yields order parameter
and energy profiles which are the same (apart from numerical
values) as those obtained by the strong streching theory of
Walton {\it et. al.} and Turner \cite{W-RMM94,turnerPRL92}.

In experiment with neutral walls, perpendicular lamellae are
always favored over unfrustrated parallel lamellae (of period
$d_0$) \cite{H-RMM98}, in contrast to the common strong
stretching prediction \cite{W-RMM94,matsenJCP97,turnerPRL92}. We
first compute the \\ bulk stability diagram and find it similar
to   previous intermediate and strong segregation calculations
\cite{matsenJCP97,G-M-B99}. We then show that proper account of
the surface change of the Flory parameter ($\tau_s>0$) can
explain the experimental findings, and significantly change the
stability diagram [compare Fig.~8(a) to 8(b)]. Thus,
perpendicular lamellae are expected to have the lowest free
energy at all separations $2L$, as in Fig.~8(b). We point out
that if the surface ordering temperature is higher than the bulk
ODT temperature, i.e. $\tau_s<0$, the $L_\parallel$ phase may
become stable even for neutral walls at $2L=nd_0$. However, this
is yet to be confirmed experimentally.

The stability diagram in this paper is similar to the diagram in
\cite{matsenJCP97}. For symmetric walls, $\sigma^+=\sigma^-$, the
$L_\perp$ is found to be stable for larger $\sigma$ fields than
the bulk approximation predicts, while for $2L\approx d_0$ it is
stable for smaller $\sigma$ fields.

One possible way to refine the calculation presented here is to
use a more accurate ansatz for the bulk order parameter
$\phi_b({\bf r})$. Such ansatz will include more $q$-modes or an
amplitude other than $\phi_q$, further lowering the free energy.

\section*{Acknowledgments}
 Partial support from the U.S.-Israel Binational
Foundation (B.S.F.) under grant No. 98-00429 and the Israel Science
Foundation founded by the Israel Academy of Sciences and Humanities
--- centers of Excellence Program is gratefully acknowledged.

%


\begin{thebibliography}{}
\bibitem{remark} For a BCP at $y>0$ confined by a surface
at $y=0$ with a uniform interaction parameter $\sigma$, the
correction field $w(y)$ is $w(y)=A_we^{-k_wy}+c.~c.~$. The
solution of Eq. (\ref{gov_perw}) subject to boundary conditions
analogous to Eqs. (\ref{bcs_perw12}), (\ref{bcs_perw34}) and
(\ref{bcs_perw5}) then gives the amplitude\\
$A_w=-\sigma k_w(4i\tau_s
k''_w+2iq_0^2(-\tau/h)^{-1/2})^{-1}$, which diverges at the ODT unless
$\tau_s\neq 0$.
\end{thebibliography}
\end{document}